  \documentclass{gretsi}
 \usepackage[english,french]{babel}   % "babel.sty" + "french.sty"
  \usepackage{times}			% ajout times le 30 mai 2003
 
 \usepackage[T1]{fontenc}
\usepackage{amsmath, amsfonts}
\usepackage{graphics}
\usepackage{array}

\titre{Analyse Multifractale des données physiologiques de marathoniens}

\auteur{\coord{Guillaume}{Saës}{1,2},
        \coord{Wejdene}{Ben Nasr}{2},
    \coord{Stéphane}{Jaffard}{2},
    \coord{Florent}{Palacin}{3},
     \coord{Véronique}{Billat}{2,3,4}}

\adresse{\affil{1}{Département
        de Mathématique, Université de Mons, Place du Parc 20, 7000 Mons (Belgique)},
        \affil{2}{Univ Paris Est Creteil, Univ Gustave Eiffel CNRS, LAMA
        UMR8050, F-94010 Creteil, France},
        \affil{3}{Laboratoire de neurophysiologie et de biomécanique du mouvement, \\ 
        Institut des neurosciences de l'Université Libre de Bruxelles, Belgique},
        \affil{4}{Université Paris-Saclay, Univ Evry, F-91000 Evry-Courcouronnes, France}}

\email{guillaume.saes@u-pec.fr,
wejdene.nasr-ben-hadj-amor@u-pec.fr,
jaffard@u-pec.fr,\\
palacinflorent@gmail.com, veronique.billat@billatraining.com
}
\resumefrancais{Nous proposons  une analyse de la fréquence cardiaque de marathoniens, mise en oeuvre par le calcul de spectres multifractals  basés sur les $p$-exposants; nous en déduisons  des conclusions physiologiques sur leurs performances. Cette analyse permet de mettre  en évidence les perturbations de l'autorégulation de la fréquence cardiaque au cours du marathon, ce qui n'a été observé pour le moment que par des échelles de ressenti des marathoniens au cours de la course.}

\resumeanglais{We propose an analysis of heart rate marathon runners  implemented by computing a multifractal spectrum based on $p$-exponents. We  draw physiological conclusions about their performance. Finally, we link this analysis with the disturbances of the heart rate autoregulation during the marathon, which had been put in evidence up to now only by scales of feeling of  marathon runners during the race.}

\begin{document}
\maketitle

\section{Introduction}
%\subsection{La classe \texttt{gretsi}}

Au cours de l'évolution, la physiologie humaine a été optimisée pour couvrir de grandes distances  afin de trouver suffisamment de nourriture pour soutenir le métabolisme du cerveau. %\cite{bramble04a, lieberman07a}
La popularité  du marathon chez les humains   est un héritage de notre capacité à courir de longues distances  en utilisant le métabolisme aérobie. % \cite{lieberman07a}. 
Le nombre de participants au marathon de Londres est passé de 7 000 à 35 000 au cours des 30 dernières années et la participation aux courses sur route en général a augmenté de plus de 50\% au cours de la dernière décennie. %\cite{pedoe07a, roberts05a, thompson09a}. 
Cette popularité  est caractérisée par l'émergence de marathoniens amateurs. 
%qui terminent l'épreuve de 42,195 km dans un un temps compris entre 2h40min et 4h40 min. 
Ils disposent désormais de cardiofréquence mètres GPS et cherchent à courir dans une zone de vitesse ou de fréquence cardiaque (FC)
% en fonction des recommandations des entraîneurs 
sans bases théoriques %physiologiques sur le bien fondé de leur choix 
et subissent une chute drastique de leur vitesse  à partir du 25ème km. %(Billat et al., 2020). 
D'autres coureurs préfèrent courir "à la sensation" et un défi important est d'analyser et d'interpréter  les séries temporelles   de leurs réponses physiologiques pour les aider à améliorer leurs performances. 
% et mécaniques 

A partir des années 1990, de nombreux auteurs ont mis en évidence le comportement fractal de données physiologiques
% de la fréquence cardiaque, la pression artérielle et la fréquence respiratoire d'êtres humains
\cite{abry10a, goldberger91a}.
En 2005, 
%E. Wesfreid, V.L. Billat et Y. Meyer
 une première analyse multifractale sur la FC de marathoniens a été  effectuée avec la  MMTO (méthode des maxima de la transformée en ondelettes) \cite{wesfreid05a}.
%, cf \cite{ABM}; 
%ces premiers résultats ne permettaient pas de mettre en évidence une évolution des paramètres multifractals au cours de la course. 
Cette étude fût complétée en 2009  en utilisant deux autres méthodes d'analyse multifractale, basées sur les  DFA (Detrended Fluctuation Analysis) %\cite{Kantelhardt}
et les coefficients d'ondelettes dominants %\cite{jaffard04a},  
appliquées sur des primitives des signaux \cite{wesfreid09a}. 
% Les limitations méthodologiques de ces approches sont dues au fait que
Les comparaisons entre ces résultats sont difficiles; en effet ces méthodes ne conduisent pas à des analyses  associées au même exposant de régularité: la MMTO est adaptée au {\sl weak scaling exponent} \cite{Meyer}, la DFA au 2-exposant \cite{pLeaders}, et les coefficients dominants à l'exposant de Hölder \cite{jaffard04a}. De plus on ne dispose de résultats mathématiques généraux  de validité que pour la méthode des coefficients dominants \cite{jaffard04a}. 
La comparaison de l'analyse de la FC  dans la première moitié et le dernier quart du semi-marathon, lorsque le coureur a commencé à avoir un épuisement du glycogène (cause majeure de la diminution de la vitesse à la fin de la course) met en évidence 
% le changement dans l'échelle fractale des fluctuations de la FC et de la vitesse. En effet, Weisfreid et al., (2009) avaient mis en évidence
un  changement  dans la dynamique de la FC
%de la vitesse et du coût de l'énergie cardiaque (le rapport fréquence cardiaque/vitesse) des coureurs
 \cite{wesfreid09a}.
%en montrant qu'il existe deux exposants d'échelle distincts.
%l'exposant à court terme et l'exposant à long terme ($\alpha_1$, pour $n$ inférieur à $n_0$ ($n_0 = 64 = 5,3$min) et la longue portée $\alpha_2$, pour $n$ supérieur à $n_0$. 
Cette étude a révélé que la fatigue 
%diminue la vitesse de course et
affecte les propriétés de régularité du signal,
%dans les exposants d'échelle à courte portée (sur de petites fenêtres) qui peuvent être considérés comme un exposant de régularité. La détection de ces changements en temps réel reste un travail à faire. 
ainsi qu'une diminution constante des rapports entre la vitesse, la cadence de pas, les réponses cardiorespiratoires (fréquence respiratoire, cardiaque, volume d’oxygène consommé) et le niveau de taux de perception de l'épuisement (Rate of Perception of Exhaustion (RPE)), selon l’échelle psychophysiologique de Borg. %Le coureur ne ressent pas la dérive de sa fréquence cardiaque, mais surtout de sa fréquence respiratoire (Billat, données personnelle). Ces données physiologiques sont peu accessibles au grand nombre et seules, les fréquences cardiaque et la cadence de pas sont des mesures disponibles pour l'ensemble des coureurs pour des raisons économiques. De plus, ces données sont générées battement par battement cardiaque et pas à pas. 

Nous confirmons et complétons ces études en montrant qu'une analyse multifractale basée sur les $p$-exposants  permet de travailler directement sur les données (et non sur des primitives) et caractérise plus précisément les modifications physiologiques durant un marathon. 
%l'objectif étant de mettre en évidence l'impact 
%conséquences de l'effort intense après le 20ème kilomètre.
%Pour cela, nous allons mettre en oeuvre une méthode d'analyse multifractale  basée sur les $p$-leaders, qui, dans certains cas, 

\section{Analyse  avec les $p$-leaders}
\subsection{Exposants de régularités ponctuelles}
 Soient $X\in L^{\infty}_{loc}(\mathbb{R})$, $t_0 \in \mathbb{R}$ et $\alpha>0$; $X$ appartient à $C^\alpha (t_0)$ s'il existe un polynôme $P_{t_0}$  et $K,r>0$ tels que $\forall t\in [t_0-r,t_0+r]$, $|X(t)-P_{t_0}(t-t_0)| \leq K |t-t_0|^\alpha.$ L'exposant  de Hölder ponctuel  est  $h_X (t_0)=\sup\{\alpha \ : \ X\in C^\alpha (t_0)\}$.
Cet exposant n'est pas défini si $X$ n'est pas localement borné;  on utilise alors la notion suivante adaptée au cadre $L^p$ \cite{jaffard05a}.
 Soient $p\in [1,+\infty[$, $X\in L^{p}_{loc}(\mathbb{R})$, $t_0\in \mathbb{R}$  et $\alpha\geq -1/p$; $X\in T^{p}_\alpha (t_0)$ s'il existe un polynôme $P_{t_0}$ , $K, R>0$  tels que 
\begin{equation*}\forall r \leq R, \; 
\left(\frac{1}{r} \int_{t_0-r}^{t_0+r}|X(t)-P_{t_0}(t-t_0)|^p dt\right)^{\frac{1}{p}} \leq K r^\alpha .
\end{equation*}
Le $p$-exposant de $X$  est 
%\begin{equation*}
$h^{(p)}_X (t_0)=\sup\{\alpha :\ X\in T^{p}_\alpha (t_0)\} .$
%\end{equation*}
Lorsque $p=+\infty$, on retrouve l'exposant de Hölder. Cet exposants de régularité se  caractérise par les coefficients d'ondelettes de la manière suivante.

\subsection{Coefficients dominants et $p$-leaders}

Soit $\psi$ une "ondelette", c'est-à-dire une fonction régulière et bien localisée telle que les $\{\psi_{j,k} (t) = 2^{j/2} \psi (2^j t-k)\}_{(j,k)\in\mathbb{Z}^2}$ forment une base orthonormée de $L^2 (\mathbb{R})$. Les coefficients d'ondelette de $X$ sont définis par $c_{j,k}=2^{j/2}\int_{\mathbb{R}} X(t) \psi_{j,k}(t) dt$.  Ils permettent de déterminer la valeur de l'{\sl exposant de régularité uniforme} $H_{min}$ de $X$ qui est caractérisé par regression $\log$-$\log$
\begin{equation}  \label{defhmain} \sup_{k} |c_{j,k}|\sim 2^{-H_{min}j} \end{equation} 
(dans la limite des petites échelles, c'est-à-dire quand $j \rightarrow + \infty$).
Si  $H_{min}>0$, alors $f$ est localement bornée et on peut déterminer l'exposant  de Hölder ponctuel de $X$  de la façon suivante:
 % Soit $X\in L_{loc}^{p}(\mathbb{R})$. 
 On note un intervalle dyadique $\lambda=\lambda_{j,k}=[k2^{-j}, (k+1)2^{-j}[$ et la réunion avec ses deux voisins est  $3\lambda=[(k-1)2^{-j}, (k+2)2^{-j}[$. Les coefficients dominants de $X$ sont définis % pour  $(j,k)\in\mathbb{Z}^2$
par
$l_{j,k}= \sup_{\lambda' \subset 3\lambda} |c_{\lambda'}|.$
Ils permettent de caractériser les exposants  de Hölder ponctuels si $H_{min}>0$ puisque pour $j\rightarrow + \infty$, on a (pour les  $\lambda_{j,k}$ contenant $x$) $l_{j,k} \sim 2^{-j h_X (x)}$, cf. \cite{abry15a}.
Si $H_{min} <  0$, l'exposant de Hölder de $X$ n'est plus défini,  mais, il est possible de travailler sur une   intégrée fractionnaires de $X$ d'ordre $\gamma >-H_{min}$.
 Pour éviter une telle transformation, on peut  utiliser le   $p$-exposant pour des valeurs de $p$ telles que $X\in L^{p} (\mathbb{R})$. Pour déterminer la valeur d'un tel $p$, il suffit de vérifier que $\eta (p) >0$ où la  \textit{fonction d'échelle ondelette}  $\eta $  est définie par $2^{-j} \sum_{k} |c_{j,k}|^p\sim 2^{-\eta (p) j}$.
%Pour une valeur $\gamma> |\lfloor H_{min} \rfloor |$, la valeur de $H_{min}$ se retrouve décalée de $\gamma$. Ainsi, $H_{min}$ est positive et il devient possible d'utiliser les leaders comme quantités multirésolution au calcul de spectre \cite{abry15a} \cite{leonarduzzi14a}. Notons qu'en pratique, on n'effectue pas une intégrée fractionnaire sur le signal, mais on multiplie les coefficients d'ondelettes par $2^{-\gamma j}$. Cette pseudo-intégrée fractionnaire présente l'avantage d'être beaucoup plus simple à effectuer tout en préservant les mêmes caractéristiques multifractales qu'une intégrée fractionnaire.
Soit $X\in L_{loc}^{p}(\mathbb{R})$. Les $p$-leaders sont définis  par
$l_{j,k}^{(p)} = \left( \sum_{\lambda' \subset 3\lambda} |c_{\lambda'}|^p 2^{j-j'} \right)^{\frac{1}{p}}.$
Ils permettent de caractériser les $p$-exposants   si $\eta (p)>0$; on a: $l_{j,k}^{(p)} \sim 2^{-j h_X^{(p)} (k2^{-j})}$ \cite{abry15a, jaffard05a}.

\subsection{$p$-spectre multifractal}
L'analyse multifractale a pour objet l'étude de  signaux dont l'exposant de régularité ponctuelle varie fortement d'un point à un autre.  Le $p$-spectre multifractal d'une fonction $X\in L_{loc}^{p} (\mathbb{R})$, est 
$D_X^{(p)} : H \mapsto \dim_{H}\left( \left\{ x\in\mathbb{R}\ : \ h_X^{(p)} (x)=H \right\} \right), $ où $\dim_{H}$ désigne la dimension de Hausdorff. 
Il est en général impossible de calculer point par point les $p$-exposants de données expérimentales. On estime plutôt  le $p$-spectre  de la façon suivante. Soit 
$S_X^{(p)} (j,q) = 2^{-j} \sum_{k} \left( l^{(p)}_{j,k}\right)^q.$
Si $S_p (j,q)\sim 2^{-j \zeta_X^{(p)} (q)}$, on appelle $\zeta_X^{(p)} (q)$ la $p$-fonction d'échelle. Sa transformée de Legendre   $\mathcal{L}_X^{(p)} (H)=\inf_{q\in\mathbb{R}} \{1+qH-\zeta_X^{(p)} (q)\}$  permet d'estimer le $p$-spectre. En effet $D_X^{(p)} (H) \leq L_X^{(p)} (H)$ \cite{jaffard04a,jaffard05a}.

\section{Analyse multifractale de la fréquence cardiaque}

\begin{figure}[htb]
    \begin{center}
    \resizebox{80mm}{!}{
    \includegraphics{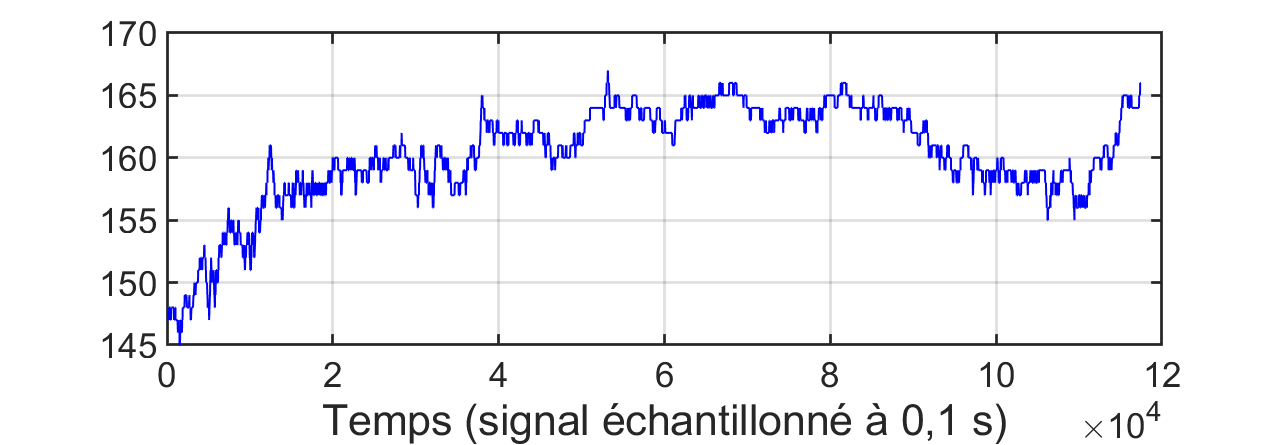}}
    \end{center}
    \legende{Fréquence cardiaque  d'un marathonien.}\label{Fig1}
\end{figure}

Nous analysons la fréquences cardiaque de 8 marathoniens (des hommes de la même tranche d'âge), cf. la Fig. \ref{Fig1}. Nous avons également considéré la cadence et l'accélération, mais elles ne permettent pas de mettre en évidence des paramètres  pertinents pour la classification;  cela peut être dû au fait que, contrairement à la cadence ou l'accélération, le coureur ne con\-trôle pas directement son rythme cardiaque.  
%la cadences (nombre de pas par minute) et l'accélération, 

\subsection{ Estimation du paramètre $H_{min}$}
\begin{figure}[htb]
    \begin{center}
    \resizebox{50mm}{!}{
    \includegraphics{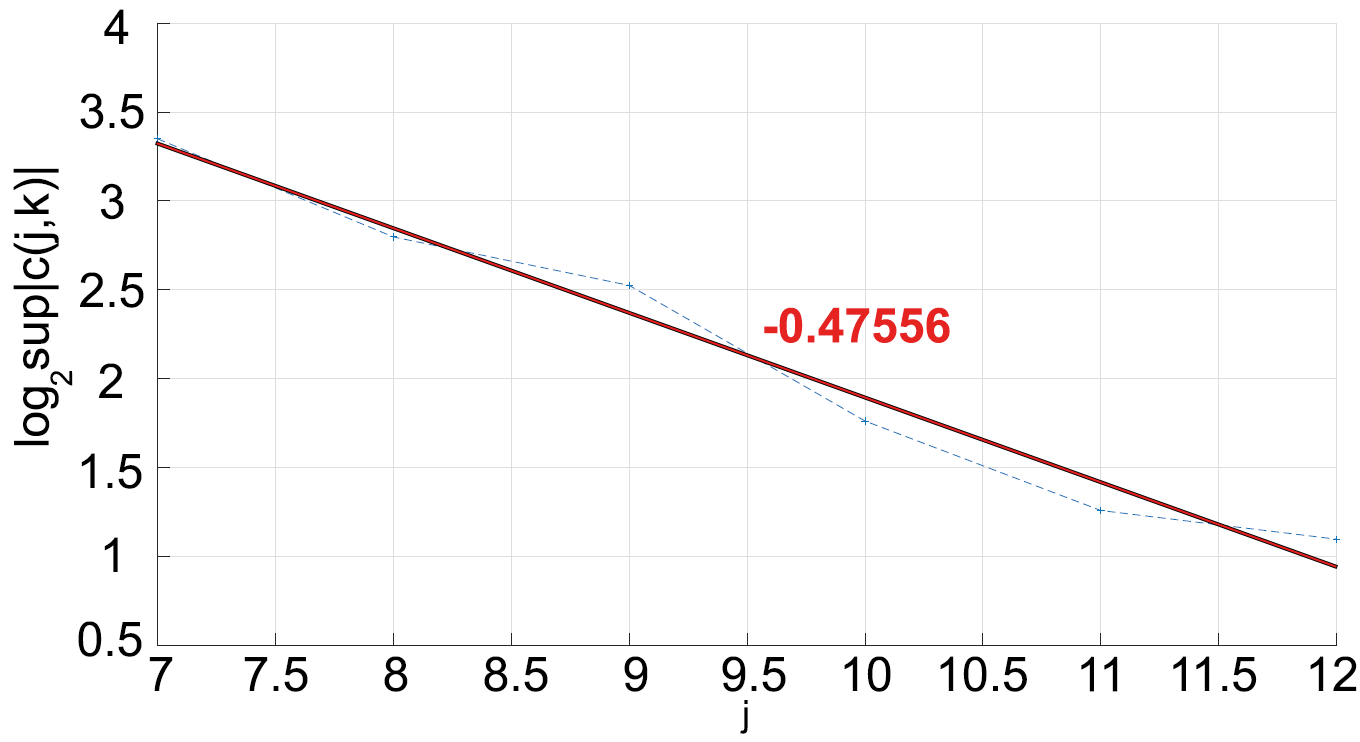}}
    \end{center}
    \legende{Estimation par régression log-log du $H_{min}$ d'une fréquence cardiaque. Le bon alignement des points montre que la régression linéaire permet d'estimer avec précision  la valeur de $H_{min}$, et  qu'il est négatif.}\label{Fig2}
\end{figure}

Le  paramètre $H_{min}$, cf. \eqref{defhmain}, permet de déterminer si une analyse  basée sur les coefficients dominants est possible, et nous verrons qu'il est également un paramètre de classification important. La Fig. \ref{Fig2} est un  exemple d'estimations de cette valeur: la régression est effectuée entre $j=8$ et $j=11$ (soit environ entre $26$s et $3$min$25$s),   échelles identifiées comme  pertinentes pour  les données physiologiques, cf.  \cite{catrambone20a}.

\begin{figure}[htb]
    \begin{center}
    \resizebox{50mm}{!}{
    \includegraphics{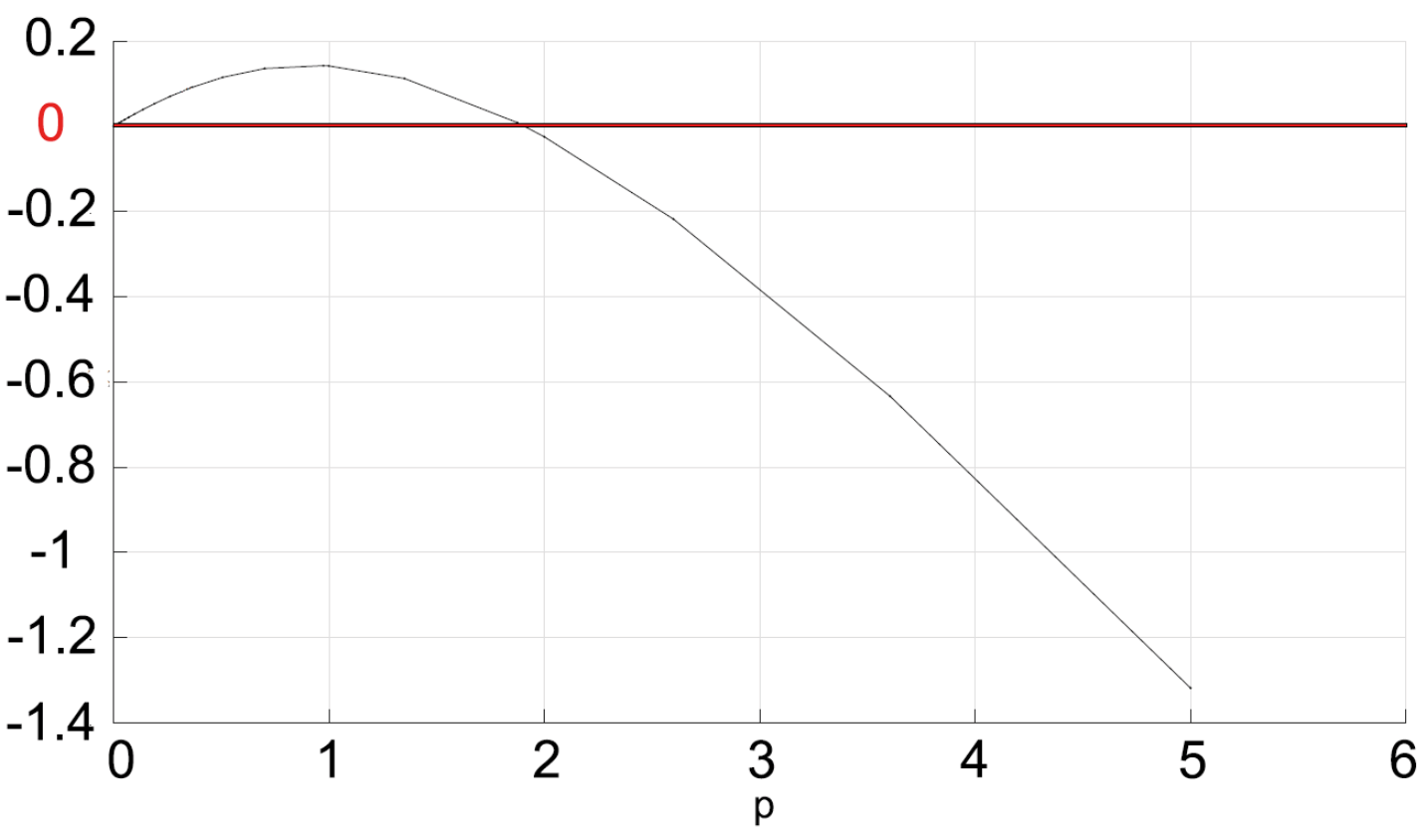}}
    \end{center}
    \legende{Exemple de fonction d'échelle ondelette de la fréquence cardiaque. Elle permet de déterminer les $p$ tels que $\zeta(p)>0$. L'estimation du $p$ spectre est alors possible.}\label{Fig3}
\end{figure}

Pour la plupart des  marathoniens, on obtient que  $H_{min}<0$ (cf. la Table \ref{tab1}) ce   qui justifie l'utilisation des $p$-leaders.
Il est  alors nécessaire de déterminer pour quelle valeurs de $p$ on a $\zeta(p)>0$, cf.  les Fig. \ref{Fig3} et \ref{Fig4}. Suite à cette première analyse, nous allons effectuer une analyse basée sur les $p$-leaders de la fréquence cardiaque.

\subsection{Détermination des 1-spectres}
\begin{figure}[htb]
    \begin{center}
    \resizebox{50mm}{!}{
    \includegraphics{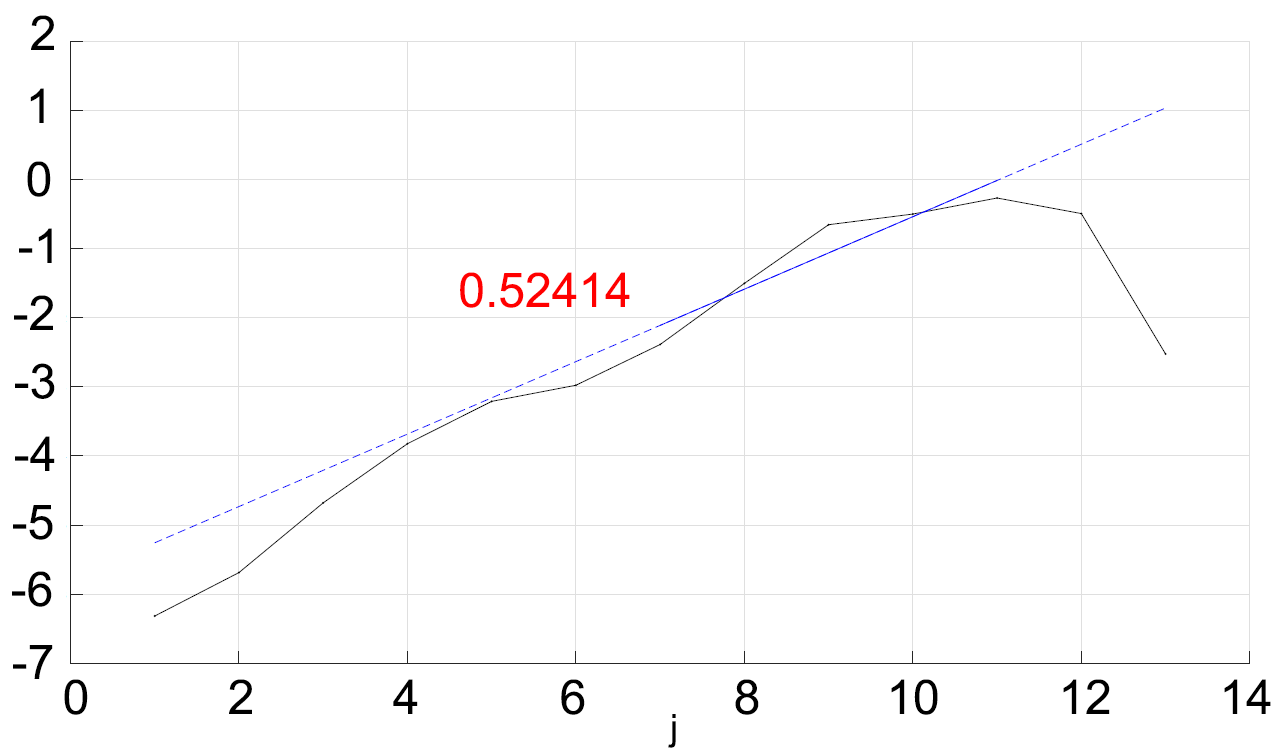}}
    \end{center}
   \legende{Estimation  par régression log-log de la fonction d'échelle ondelettes de la fréquence cardiaque pour $p=1$. La  pente de la régression  est positive. L'utilisation de  $1$-leaders pour l'analyse multifractale est ainsi justifiée.}\label{Fig4}
\end{figure}

\begin{figure}[htb]
    \begin{center}
    \resizebox{70mm}{!}{
    \includegraphics{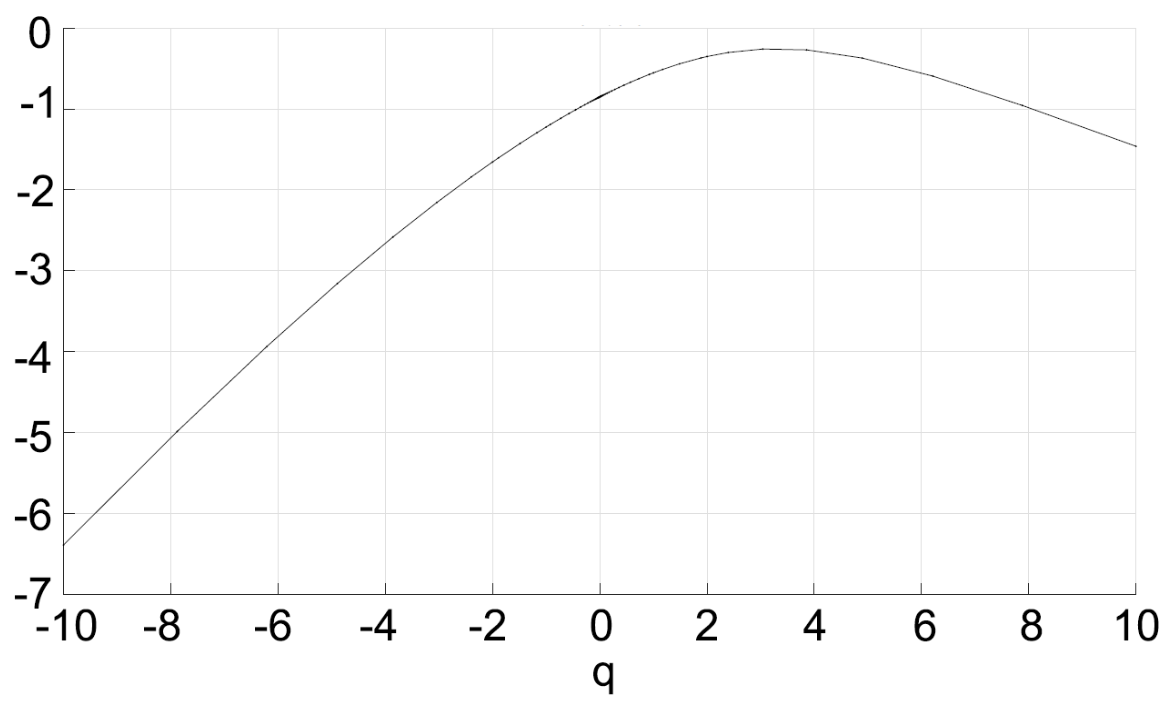}
    \includegraphics{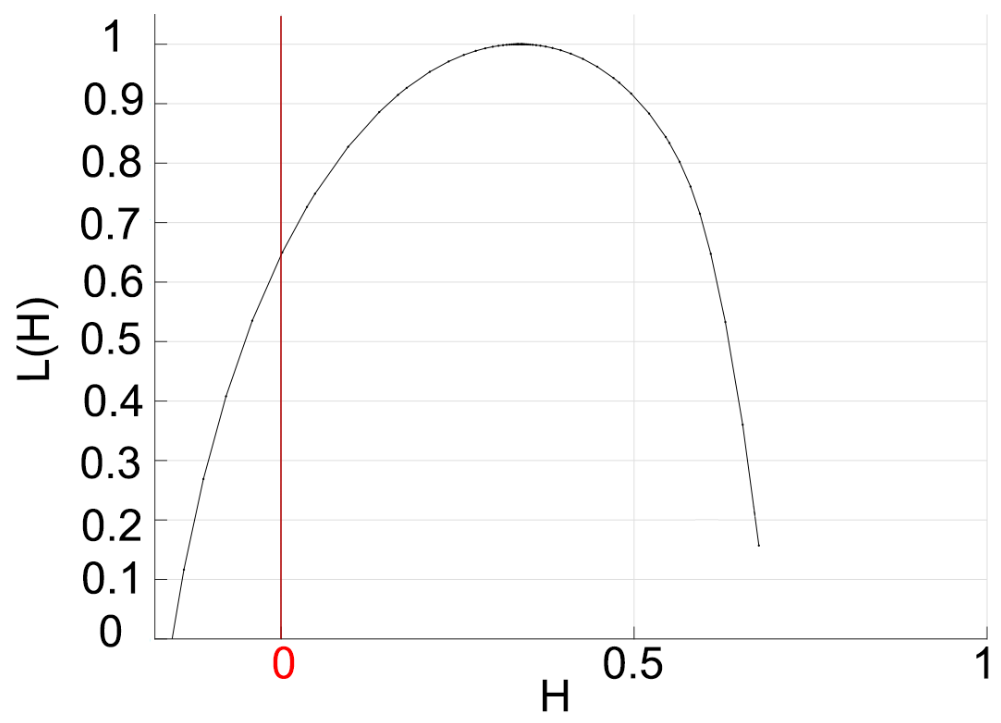}}
    \end{center}
    \legende{Estimations de la fonction d'échelle 1-leaders de la fréquence cardiaque (à gauche) et de sa transformée de Legendre (à droite).  La partie gauche du spectre débute pour des exposants $H<0$ confirmant ainsi la présence de singularités d'exposant négatif   dans le signal.}\label{Fig5}
\end{figure}

Nous avons choisi de prendre $p=1$  et $p= 1,4$ pour le calcul des  $p$-leaders  car les données étudiées vérifient toujours $\zeta(p)>0$ pour ces valeurs, cf. la Fig. \ref{Fig3}. La Fig. \ref{Fig4} fournit un tel exemple  d'estimation de  $\eta (1)$.
Il est  alors justifié d'estimer le $1$-spectre multifractal pour les données physiologiques sur la fréquence cardiaque des marathoniens. Sur la Fig. \ref{Fig5}, on représente un exemple de $1$-spectre de Legendre.

On effectue alors une classification des données sur les 8 marathoniens avec des valeurs $p=1$ et $p=1,4$ (cf. Table \ref{tab1}). 
%Nous donnons aussi les résultats pour $p=2$ mais ils ne sont pas pertinent pour le dernière marathonien M2 puisque $\zeta(2)<0$. 
%Ce ratio entre $H_{min}/c_1$ permet de révéler des différences d'évolution des paramètres multifractaux au cours du marathon en fonction des types de coureurs (expérience du marathon, âge, niveau de performance) (cf. Fig. \ref{Fig6}).

\begin{table}[htb]
    \legende{Analyse multifractale de la fréquence cardiaque des marathoniens ($\widetilde {H}_{min}$ et $\widetilde{c}_1^p$ sont respectivement le $H_{min}$ et $c_1^p$ de la primitive du signal)}\label{tab1}
    \begin{center}
    \resizebox{88mm}{!}{
    \begin{tabular}{||c||*{10}{m{1.2cm}|}|}
        \hline\hline
        $ $   & $H_{min}$  & $\widetilde{H}_{min}$ & $c_1^1$&  $c_1^{1,4}$   & $\widetilde{c}_1^1$ & $\widetilde{c}_1^{1.4}$  \\ 
        \hline
        M1 & $-0,2768$ & $0,7232$ & $0,8099$ & $0,8064$ & $1,8242$ & $1,8213$  \\
        \hline
        M2 & $-0,0063$ & $0,9937$ & $0,4564$ & $0,4043$ & $1,3926$ & $1,3509$  \\
        \hline
        M3 & $-0,0039$ & $0,9961$ & $0,6856$ & $0,6625$ & $1,6942$ & $1,6351$ \\
        \hline
        M4 & $-0,1633$ & $0,8367$ & $0,6938$ & $0,6785$ & $1,6653$ & $1,6636$ \\
        \hline
        M5 & $-0,2434$ & $0,7566$ & $0,5835$ & $0,5689$ & $1,5401$ & $1,5224$ \\
        \hline
        M6 & $-0,3296$ & $0,6704$ & $0,5809$ & $0,5636$ & $1,5644$ & $1,5500$  \\
        \hline
        M7 & $0,1099$ & $1,1099$ & $0,5652$ & $0,5483$ & $1,4754$ & $1,4379$  \\
        \hline
        M8 & $-0,5380$ & $0,4620$ & $0,3382$ & $0,2977$ & $1,2588$ & $1,2086$  \\
        \hline\hline
    \end{tabular}}
    \end{center}
\end{table}

\begin{figure}[htb]
    \begin{center}
    \resizebox{48mm}{!}{
    \includegraphics{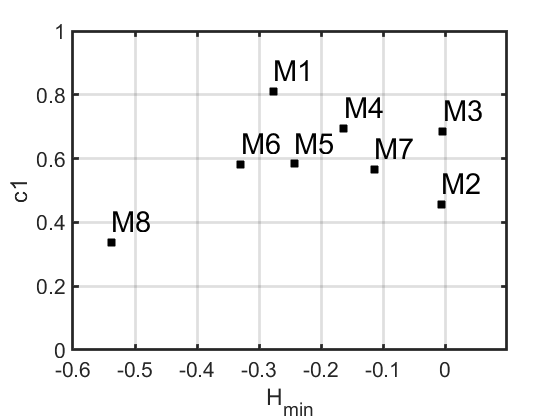}}
    \end{center}
    \legende{Représentation du couple $(H_{min}, c_1)$ déduits du 1-spectre de la fréquence cardiaque; $H_{min}$ apparaît comme le paramètre de classification  le plus pertinent. Le point isolé à gauche correspond à M8, le coureur le plus entraîné.}\label{Fig6}
\end{figure}

Dans la Fig. \ref{Fig6}, on représente pour chaque marathonien les valeur $(H_{min}, c_1^1)$ où $c_1^p$ est la valeur de  $H$  pour laquelle le maximum du $p$-spectre est atteint. Il  correspond mathématiquement à l'exposant presque partout du signal. 
Les valeurs de $c_1^1$ restent très proches de $0,4$ tandis que celles du $H_{min}$ varient fortement,  mettant  en évidence des différences entre coureurs plus ou moins expérimentés. Les autres paramètres permettant de caractériser la forme des spectres ($c_2^p$ et $c_3^p$ \cite{abry10a}) ne s'avèrent pas pertinents pour la classification, ainsi nous ne les présenterons pas. Le coureur M8 qui présente le $H_{min}$ le plus petit est le seul coureur de \textit{trail}, et il a surpassé son précédent record; de plus, il est le plus entraîné  avec une façon de courir (nombre de pas par minute et vitesse/accélération) très irrégulière. 
%A l'opposé,  M7 est le coureur le moins entraîné avec un régime de course très lent et détendu, et  sa façon de courir est très régulière. 

La Table \ref{tab1} montre que les valeurs  de $c_1^p$ varient très peu en fonction de $p$,  et que, pour des primitives, le décalage des paramètres $H_{min}$ et $c_1^p$ est d'environ 1, ce qui est caractéristique de signaux ne contenant que des singularités de type { \em cusp} (absence de chirps dans le signal), et montre le caractère intrinsèque de la valeur de $c_1^p$ pour ce type de données. Ces résultats sont confirmés en vérifiant que le spectre d'une primitive du signal se translate de $1$.
%Ceci permet de confirmer qu'il n'y a pas la présence de singularité oscillante sur le signal de fréquence cardiaque. 

\subsection{Classification des phases d'un marathon}

\begin{figure}[htb]
    \begin{center}
    \resizebox{77mm}{!}{
    \includegraphics{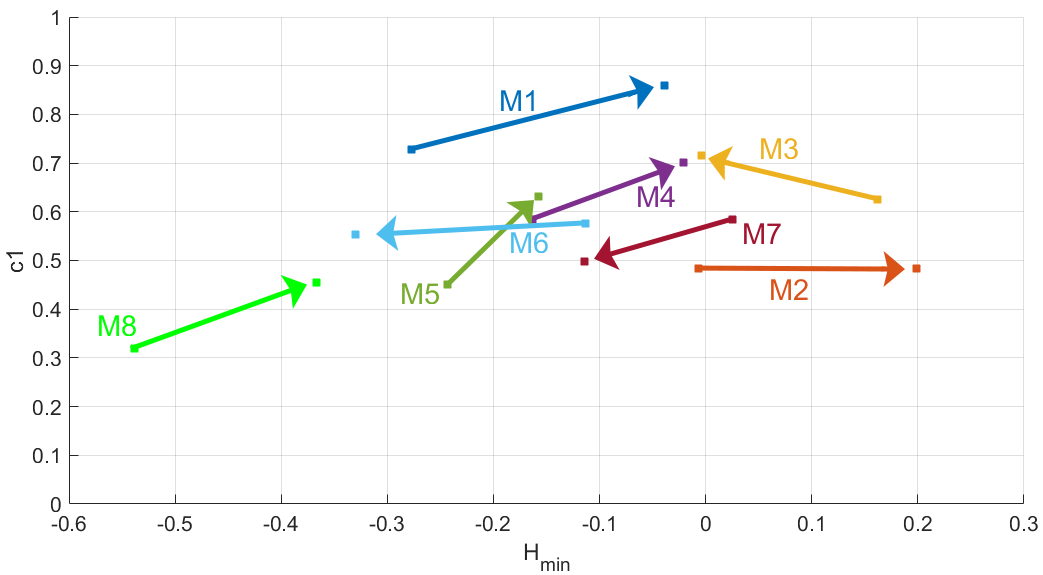}}
    \end{center}
    \legende{Évolution de $(H_{min}, c_1^1)$ déduits du 1-spectre de la fréquence cardiaque entre le début (en bleu) et le dernier quart (en rouge) du marathon : les évolutions sont similaires sauf pour trois coureurs : M3 et M6 qui ont éprouvé de grandes difficultés et  M7 qui est le moins expérimenté  avec  un temps de course nettement plus long et une façon de courir très régulière.}\label{Fig8}
\end{figure}

Nous nous intéressons aux évolutions des paramètres multifractals lors du marathon. Vers le 25ème kilomètre (environ à $60\%$) les  coureurs ressentent une  pénibilité accrue sur l'échelle RPE Borg (Rate of Perceived Exertion)  utilisée pour identifier l'intensité d'un exercice en fonction du ressenti. %Ce changement  apparait pour les 8 marathoniens  entre le 26ème et 30ème kilomètre. 
La Fig. \ref{Fig8} permet d'observer l'évolution des paramètres de multifractalité  entre la première  moitié du marathon  et  le dernier quart,  mettant  en évidence les différences de réactions physiologiques face à la fatigue ressenties à partir du  28ème kilomètre. 

%On , sauf pour M2. L'interprétation physiologique possible sur ces résultats est la fatigue et la gêne que les différents marathoniens ont commencé à rencontrer au bout du 30ème kilomètre de la course. Par ailleurs, les évolutions se font presque toutes de la même manière (vers le bas gauche) sauf pour M7 qui est le coureur le moins expérimenté de la course avec u un temps de course nettement plus long.

\section{Conclusion et perspectives}

L'analyse au moyen de $p$-leaders du rythme cardiaque de marathoniens permet de compléter les analyses précédentes basées sur la MMTO, la DFA, ou les coefficients dominants. Sur le plan méthodologique, nous avons mis en évidence le fait qu'une analyse basée sur les 1-exposants permet de travailler directement sur les données et non sur des primitives. De plus  l'analyse des différentes phases de la course permet de relier l'évolution des paramètres multifractals et la physiologie des coureurs : L'exposant $H_{min}$ évolue peu durant la course  pour des coureurs expérimentés, mais diminue en fin de course pour les autres, ce que l'on peut interpréter comme une désorganisation dans leurs efforts. De plus, ces résultats confirment des analyses réalisées sur des matrices physiologiques complètes (cardio respiratoire), nécessitant des mesures coûteuses avec des analyseurs ne permettant pas de restituer des données en qualité suffisante comme le fait le cardiofréquence mètre mesurant chaque période cardiaque, avec une précision de 0.001s soit 0.01 hertz pour la fréquence cardiaque). Ces cardiofréquence mètres sont à présent largement utilisés par les coureurs, en revanche, le traitement des série temporelles post course se fait tout au plus avec la méthode DFA.

Nous avons mentionné que  l'analyse multifractale   de  la cadence et de  l'accélération des coureurs  ne fournit pas de  paramètres  pertinents pour la classification. Il est cependant possible qu'une analyse multivariée qui utilise conjointement ces données avec la fréquence cardiaque (en suivant les outils développés dans \cite{Seuret}) puisse fournir des informations plus riches. Ce point sera l'objet d'une étude future.

\end{document}